\documentclass[aps,reprint,prx,nofootinbib,showkeys]{revtex4-1}
\usepackage{amsfonts, amsmath, amssymb, wasysym}
\usepackage[pdftex]{graphicx}
\usepackage{natbib}

\newcommand{\CC}{{\mathcal C}}

\newcommand{\EE}{{\mathcal E}}
\newcommand{\RR}{{\mathcal R}}
\newcommand{\MM}{{\mathcal M}}
\newcommand{\NN}{{\mathcal N}}

\newcommand{\vsigma}{{\vec\sigma}}

\newcommand{\mytitle}
    {Theoretical ecology without species}
\newcommand{\myauthor}
    {Mikhail Tikhonov}
\newcommand{\myaffilA}
    {Center of Mathematical Sciences and Applications}
\newcommand{\myaffilB}
    {John A. Paulson School of Engineering and Applied Sciences}
\newcommand{\myaffilC}
    {Kavli Institute for Bionano Science and Technology, Harvard University, 29 Oxford St, Cambridge, MA 02138, USA}

\begin{document}
\title{\mytitle}
\author{\myauthor}
\email{tikhonov@fas.harvard.edu}
\affiliation{\myaffilA}
\affiliation{\myaffilB}
\affiliation{\myaffilC}
\keywords{microbial ecology; community dynamics; species; resource competition}
\begin{abstract}
Ecosystems are commonly conceptualized as networks of interacting species. However, partitioning natural diversity of organisms into discrete units is notoriously problematic, and mounting experimental evidence raises the intriguing question whether this perspective is appropriate for the microbial world. Here, an alternative formalism is proposed that does not require postulating the existence of species as fundamental ecological variables, and provides a naturally hierarchical description of community dynamics. This formalism allows approaching the ``species problem'' from the opposite direction. While the classical models treat a world of imperfectly clustered organism types as a perturbation around well-clustered ``species'', the presented approach allows gradually adding structure to a fully disordered background. The relevance of this theoretical construct for describing highly diverse natural ecosystems is discussed.
\end{abstract}
\maketitle

Although the basic unit participating in ecological interactions is an individual organism, constructing tractable theoretical models usually requires clustering individuals into discrete groups within which organisms are treated as identical; such classification can be based, for example, on taxonomy, development stage, or phenotype~\cite{May07}. The resulting tension is a long-standing issue in community ecology: ever since Darwin~\cite{Ereshefsky10}, the difficulty of drawing sharp boundaries partitioning natural diversity into discrete categories~\cite{Hey06} and the realization that the individual-level variation can be an important actor in ecological phenomena~\cite{DeAngelis05,Bolnick11} made the ``partitioned community'' assumption of well-delimited, uniform groups highly problematic~\cite{Shapiro14,DeQueiroz07,Achtman08,Konopka09,Hart11,Cordero14}.

The urgency of this issue has been highlighted over the last decade by the studies of microbial diversity in natural environments~\cite{Gill06,Turnbaugh09,Caporaso11,Lozupone12,HMP,EMP}. For microbes, the partitioning problem is intensified by the prevalence of asexual reproduction and horizontal gene transfer~\cite{Staley06,Fraser09}, and the fact that even genetically identical cells can assume different ecological roles, e.g.\ in a biofilm~\cite{Lyons15}. To cope with these issues, considerable attention has been devoted to identifying questions that can be asked and answered without explicitly specifying a partitioning~\cite{Konopka09}, e.g.\ solely in terms of community metagenome~\cite{Tringe05,Konopka09,Vieites09}. These approaches adopt a convenient coarse-grained viewpoint that is appropriate for certain functional questions, e.g. comparing communities~\cite{Tringe05}. However, their ability to describe community dynamics is necessarily limited. For that purpose, the partitioned community assumption currently has no alternatives: although conceptually problematic, it is seen as an operational necessity~\cite{Sites04,Cohan06,Cohan07,Shapiro14}. Short of resorting to individual-based modelling~\cite{DeAngelis05}, it remains unclear how the dynamics of ecological communities could be described in a naturally hierarchical way~\cite{Ereshefsky01,Konopka09,Hart11,Okasha11,Shapiro14}).

The hot debate surrounding the definition of the ``unit'' of ecological diversity prompts an intriguing question: is the partitioned community assumption an adequate description of the microbial world? The species-based intuition is incontestably useful, but might it be forcing onto our data a structure that it does not possess~\cite{Hendry00,Doolitle12}? Ultimately, of course, this question should be settled by experiments, and reports are conflicting~\cite{Acinas04,Hunt08,Shapiro12,Kashtan14,Biller15}. However, ideally we should be asking not whether a species-based picture is ``adequate'', but whether it is superior to alternatives. Can we imagine an ecology where no grouping of individuals is privileged, and how would we describe it? Regardless of our stance on the ``species problem'', we must recognize this question as an essential theoretical exercise: it is impossible to test a hypothesis unless we can imagine it to be false. Without an alternative, the partitioning assumption will continue to shape our questions, our models and even our data analysis in the form of ``operational taxonomic units'', an ill-defined proxy for a ``bacterial species'' extensively critiqued elsewhere~\cite{Schmidt15}.

\begin{figure}[b!]
\centering
\includegraphics[width=0.97\linewidth]{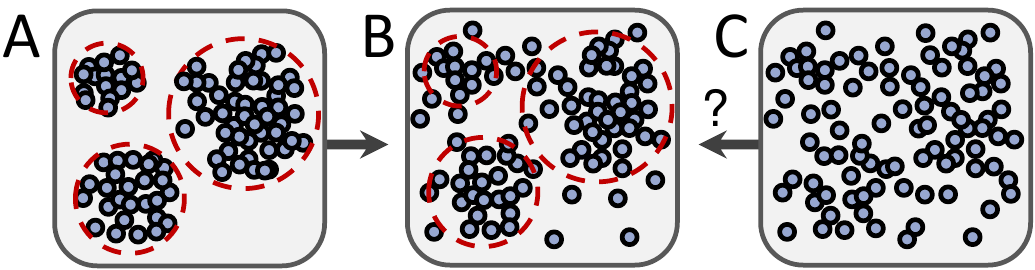}
\caption{Clustering organisms into discrete species is often problematic \textbf{(B)}. Rather than describing the natural world as a deviation from a perfectly clustered case \textbf{(A)}, this work proposes a theoretical construct where structure can be gradually added to a fully disordered ``ecology without species'' \textbf{(C)}.
\label{fig:idea}}
\end{figure}

Borrowing an idea from condensed matter physics, this work proposes a new way to tackle the species problem, approaching it ``from the other side'', as illustrated in Fig.~\ref{fig:idea}. Consider a heterogeneous community where some individuals are more alike than others, but the clustering is imperfect, as is arguably the case in most natural situations~\cite{Hey06,Fraser09,Hendry00,Schmidt15}. Currently, the only existing approach treats this scenario as a small deviation from the classic picture of well-clustered species. In physical terms, this is a perturbative expansion, where we start from a perfectly partitioned world (Fig.~\ref{fig:idea}A) and attempt to recognize that species boundaries are ``fuzzy'' (Fig.~\ref{fig:idea}B). This work proposes an explicit construct where the perturbative expansion can be performed around a different origin, starting from an unstructured phenotype background (Fig.~\ref{fig:idea}C), and adding some structure.

Similar ideas have a long history and have been enormously productive in many areas of physics (cf. the classic theory of metals that treats electrons as a free gas). The contribution of this work is to port this idea into an eco-evolutionary context. In order to achieve this, it is necessary to first develop a formalism that avoids postulating a community partitioning. This work extends the notion of an ``ecomode''~\cite{Leibler12,Leibler15} to construct a naturally hierarchical, rank-free description of community dynamics in terms of population fluctuation eigenmodes. This approach reduces to the familiar species-based perspective when it applies, but remains well-defined even when the species description breaks down. The basic idea behind this formalism is that species clustering can be reinterpreted as a special form of a more general operation, namely a change of basis in the compositional space.

\section{Reinterpreting species clustering as a change of basis}
Describing an ecosystem is a problem of dimensionality reduction. Given a community of billions of individuals, or millions of strains, the challenge is to find a reduced set of degrees of freedom that could allow characterizing its properties or dynamics with an acceptable degree of accuracy.

\begin{figure}[t!]
\centering
\includegraphics[width=0.8\linewidth]{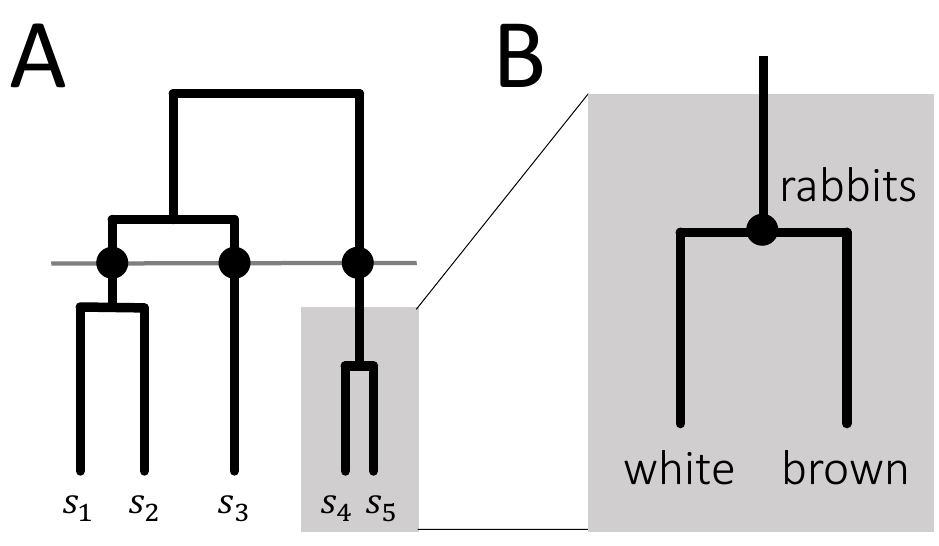}
\caption{\textbf{Coarse-graining degrees of freedom through clustering.}
\textbf{A.} The most common approach to coarse-graining the description of an ecosystem is by merging branches of a phylogenetic tree.
\textbf{B.} An elementary clustering operation where the abundances of two rabbit varieties are combined into a single degree of freedom.
\label{fig:1}}
\end{figure}

The standard approach to this problem is through clustering. Most commonly, the microscopic degrees of freedom, such as abundance of individual strains, are assumed to be located at the leaves of a tree (based on phenotypic similarity, phylogenetics, etc.). A coarse-grained description is then constructed by choosing a transect cutting across this tree, and combining branches lying downstream of the transect (Fig.~\ref{fig:1}A). For concreteness, consider an example of combining just two leaves. Imagine a community where a species of rabbits comes in two varieties that differ only slightly: say, a ``white'' rabbit $W$ and a ``brown'' rabbit $B$ (Fig.~\ref{fig:1}B). When describing their predator-prey dynamics with foxes, a two- or three-variable description could be used, depending on the desired level of detail. The comprehensive three-variable description might capture come subtle features, but for many questions, the classic Lotka-Volterra model, whereby the two varieties are clustered together into a single variable ``rabbits'' $R$, would provide a sensible approximation:
\begin{equation}\label{eq:WBR}
\{W,B\}\mapsto R.
\end{equation}

This work proposes a different perspective on the clustering operation~\eqref{eq:WBR}. To a first approximation, most environmental perturbations will affect the population of rabbits as a whole; e.g.\ an increase in predator efficiency, or habitat deterioration will depress the population of both varieties together. Upon careful examination, however, the small functional differences will manifest themselves on a slow timescale: for example, if the predator is better attuned to spotting a particular coat colour, one variety may eventually outcompete the other. In this way, compositional changes naturally align not with the Lotka-Volterra axes $W$ and $B$, but with the rotated axes that provide an hierarchically-informed description: one variable $R=W+B$ (``rabbits'') that changes on a fast timescale, and another $C=W-B$ (``coat color'') that changes on a slow timescale. This work proposes to reinterpret Fig.~\ref{fig:1}B not as clustering, but as a change of basis, followed by the decision to ignore a slow-timescale variable. As we will see, an appropriately rotated basis can be constructed for an arbitrary population of interacting individuals, and can serve as a naturally hierarchical characterization of a community. Borrowing the term introduced in~\cite{Leibler12}, one could call this the ``ecomode description''.

\section{The eigenmode description:\\ the general definition}
Let $\CC$ denote a full individual-based description of a community: a list of all present organisms (labelled by indices $\mu$, $\nu$\dots) and their functional characteristics. A general model for ecological dynamics can be written as a rule that associates to each individual $\mu$ a ``rate of abundance change'' $r_\mu$: the probability per unit time with which it will either generate another individual ($r_\mu>0$) or die ($r_\mu<0$). In a most general model, this rate can depend on any detail of the community state, as well as on external environmental parameters denoted by $\EE$:
\[
r_\mu=r_\mu(\EE,\CC)
\]
Within any such model, we can define a matrix $\MM$ of interactions between individuals: $\MM_{\mu\nu}$ is the effect that a hypothetical duplication of individual $\mu$ would have on the instantaneous division/death rate of $\nu$.
\begin{equation}\label{eq:Mindividual}
  \MM_{\mu\nu}=r_\nu(\EE,\{\text{$\CC$ with $\mu$ duplicated\}})-r_\nu(\EE,\CC)
\end{equation}
Importantly, here $\mu$ and $\nu$ are \emph{individuals}, not species, and so the definition of $\MM$ does not require the partitioned community assumption. For the classic Lotka-Volterra scenario, the ``functional characteristics'' of an individual need only specify whether it is a fox or a rabbit, but we can allow this microscopic description to be arbitrarily detailed, down to the point of all individuals being unique. Note that the interaction matrix $\MM$ is defined instantaneously; the number of individuals and thus the size of $\MM$ can change in time.

The primary focus of this work is the eigensystem of the interaction matrix $\MM$. For a population of $\NN$ individuals, this $\NN\times \NN$ matrix will always have $\NN$ eigenvalues. Note, however, that if individuals $\mu$ and $\mu'$ happen to be functionally identical, then $\mathcal M_{\mu\nu}=\mathcal M_{\mu'\nu}$ for all $\nu$, and $\mathcal M$ has a zero eigenvalue. For a coarse model ascribing all individuals to only $K\ll\NN$ distinct phenotypes, the matrix $\MM$ will be highly degenerate with only $K$ non-trivial eigenvalues; the remaining $\NN-K$ will be strictly zero. A detailed functional description recognizing all individuals as unique will remove the degeneracy; however, if the partitioning into $K$ categories were indeed a good approximation, $\NN-K$ eigenvalues will remain small (here and everywhere, characterizing an eigenvalue as ``small'' refers to its absolute value). We are led to hypothesize that the spectrum of $\MM$ may provide a naturally hierarchical description of ecological relationships in a community, with progressively smaller eigenvalues resolving finer and finer details. The grouping of individuals into discrete types, when it exists, is established as a result of this analysis; in this framework, it no longer constitutes a fundamental assumption.

\section{Restricting generality: eigenmodes in Lotka-Volterra models}
\begin{figure*}[t!]
\centering
\includegraphics[width=0.9\textwidth]{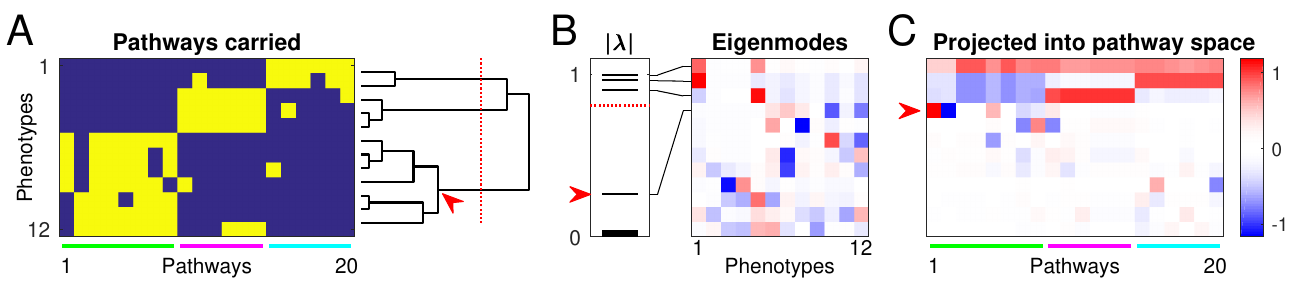}
\caption{For a well-clustered set of phenotypes, the eigenmode representation captures the hierarchy of phenotype similarity. \textbf{A:}~Pathway presence matrix for 12 phenotypes, generated as variations on 3 ``core phenotypes'' (colored lines); phenotypes carrying a given pathway (shown in yellow) are able to metabolize the respective resource.  The tree represents hierarchical clustering based on Hamming distance. Dotted transect corresponds to the clear grouping into 3 clusters. Under MacArthur's model, resource competition in this ecosystem leads to a unique stable equilibrium. \textbf{B:} The spectral structure of the ecosystem shown in A, in the vicinity of its equilibrium point. Left: the eigenvalues of the Jacobian. All eigenvalues are negative, corresponding to a stable equilibrium; shown is the absolute value. Right: the eigenvectors, ordered by decreasing $|\lambda|$; each is a linear combination of 12 phenotypes and is difficult to interpret directly. \textbf{C:} The same eigenvectors,  projected into pathway expression space. It is now clear that the three dominant modes span the three ``core phenotypes'' (colored lines as in panel A). The fourth mode (arrowhead) corresponds to a subdivision of the first cluster (compare with panel A).
\label{fig:2}}
\end{figure*}
In the interest of full generality, the definition of the previous section was explicitly individual-based. The existence of such a definition is a very important feature of the eigenmode framework; however, the inconvenience of dealing with discrete individuals is a technical nuisance that could obscure the larger points in the subsequent discussion. To build intuition about the eigenmode perspective, the best approach is to illustrate it on simple examples.

The most familiar setting is that of a generalized Lotka-Volterra model, with $K$ continuous degrees of freedom. Its linearized dynamics in the vicinity of an equilibrium point take the form:
\begin{equation}\label{eq:contDynamics}
\frac{d\,\delta n_a}{dt} = \sum_{b=1}^K M_{ab} \,\delta n_b.
\end{equation}
Here $\delta n$ denotes the deviation from equilibrium abundance, and the index $a$ runs from 1 to $K$. For simplicity, this is the context that will be used for most of the discussion. Positing a finite number of $K$ interacting phenotypes may seem to contradict the stated goal of avoiding the ``species'' partitioning. However, the reader is invited to think of these phenotypes not as a ``species'' label, but as a microscopic, sub-species description; e.g.\ interacting bacterial strains that may or may not group into species. In this context, the variable-size individual interaction matrix $\MM$ has only $K$ non-trivial eigenvalues and eigenvectors, which correspond to the eigensystem of the Lotka-Volterra interaction matrix $M$ (see the Supplementary Material; SM). The eigenvalues have units of inverse time and correspond to the relaxation time scales of perturbations applied along each eigenmode.

In order to make the eigenmodes interpretable, rather than positing the interaction matrix $M$ in some arbitrary manner, it is extremely helpful to construct it from a functional description of what these phenotypes do (recall the example above, where the $W+B$ axis could be conveniently interpreted as ``rabbits'' and $W-B$ as ``coat color''). One simple way to do this is provided by the MacArthur model of resource competition~\cite{MacArthur67,MacArthur69,MacArthur70}, where phenotypes are characterized by the resources they are able to consume. This defines a particular instance of a competitive Lotka-Volterra model, while also giving us an intuition for phenotypic similarity, and a functional interpretation for the eigenmodes we will observe.

Specifically, following ref.~\cite{CWC}, consider a well-mixed microbial community in a medium where a single limiting element (e.g.\ carbon) exists in $N$ forms (``resources'' $i\in\{1\dots N\}$; in what follows $N=20$). Each resource can be metabolized with a dedicated ``pathway''. The phenotypes are defined by two pieces of information: first, their requirement for the limiting element, and second, the pathways they carry. The former will be randomly drawn for simplicity, but the latter will be chosen strategically, by hand, in order to construct the most instructive examples. The set of $K$ phenotypes defines a $K$-by-$N$ binary matrix $\sigma$ of pathway presence/absence ($\sigma_{ai}=1$ if phenotype $a$ carries the pathway enabling it to consume resource $i$), and this matrix $\sigma$ determines the phenotype-phenotype interactions. Specifically, the competition between a pair of phenotypes $(a,b)$ is the stronger, the larger the overlap in their resource preference $\sum_i\sigma_{ai}\sigma_{bi}$. Under MacArthur's model, such competition always results in a unique and stable equilibirum~\cite{MacArthur69,CWC}. The mathematical details are provided in the SM, but for the discussion that follows, this description should already be sufficient.

\section{The effective number of species is defined by a spectral gap}
For the first example, consider a community composed of phenotypes depicted in Fig.~\ref{fig:2}A, competing for 20 resources supplied in equal abundance. These 12 phenotypes were purposefully generated to form three clear clusters (see SM for the exact procedure; \textsc{Matlab} scripts (Mathworks, Inc.)
reproducing all examples and figures are available as Supplementary file~1). This community can be interpreted as harbouring three species; within each species, phenotypes share a core set of pathways (indicated) and differ only in a small subset, similar to the ``core'' and ``accessory genome'' of a species~\cite{Medini08}. How is this scenario seen from the eigenmode perspective?

Resource competition between these phenotypes defines a 12-by-12 interaction matrix. Its eigenmode spectrum is presented in Fig.~\ref{fig:2}B. Each eigenvector is a linear combination of phenotypes, and is difficult to interpret. However, since each phenotype is defined by its resource consumption, their linear combination can be conveniently characterized in the same way: the pathway presence/absence matrix $\sigma$ can be used to project each eigenvector into the pathway expression space (Fig.~\ref{fig:2}C). This provides a functional interpretation for the eigenmodes: each row in the matrix shown in Fig.~\ref{fig:2}C indicates how the community-level pathway expression changes if its composition is perturbed in the direction of a given eigenmode.

We observe that the spectrum is dominated by 3 modes, whose projections in the pathway expression space are linear combinations of the ``core genomes'' of the three phenotype clusters (Fig.~\ref{fig:2}C, top 3 rows). This indicates that at the short timescale, the response of the system to a perturbation can be well approximated using only 3 degrees of freedom. Above, characterizing this habitat as ``harbouring 3 species'' was an interpretation that relied on a subjective judgement of Fig.~\ref{fig:2}A by a human observer. The eigenmode spectrum contains this information, and more, in an objective format, decomposing community dynamics over a range of time scales.

The first eigenvector is a collective mode, with each entry close to $1$. This corresponds to the statement that, if we insisted on reducing the community to a single degree of freedom, our best choice would be to say that all resources are being consumed at a certain overall rate. However, this is a poor approximation; capturing the dominant fast dynamics requires 3 degrees of freedom. Far below the dominating eigenvalues, a fourth one becomes apparent (arrowhead). Examining the fourth row in Fig.~\ref{fig:2}C and comparing it to panel A reveals that this mode corresponds to a finer structure within the first cluster, where coexisting variants differ in their ability to consume resources 1 and 2. The eigenmode decomposition provides a way to describe dynamics with increasing level of detail; smaller differences manifest themselves as smaller eigenvalues. Thus, the eigenmode description of community structure is naturally hierarchical.

\begin{figure*}[t!]
\centering
\includegraphics[width=0.8\textwidth]{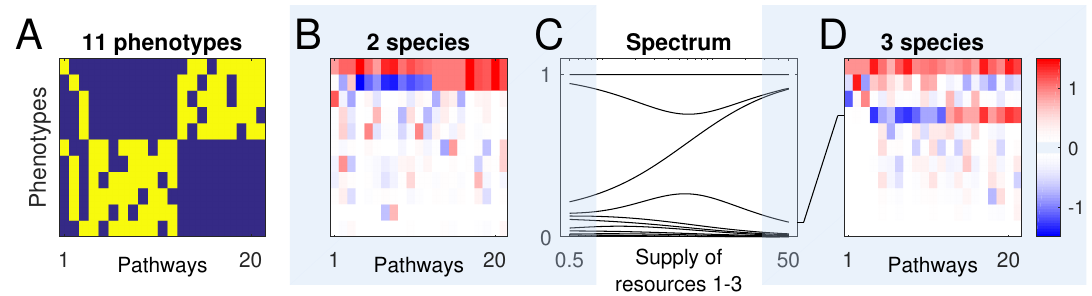}
\caption{A smooth interpolation between two unrelated clustering patterns into, respectively, two and three effective species.
\textbf{A:} Pathway presence matrix (same format as in Fig.~\ref{fig:2}A) for 11 phenotypes. Their preference for resources 1--3 induces a grouping into three clusters, but the remaining pathways induce a conflicting clustering pattern (with only two groups). \textbf{B:} If the supply of resource 1--3 is low, community dynamics exhibits two dominant modes, reflecting the 2-cluster structure. As the supply of these resources is increased (panel \textbf{C}), the preference for resources 1--3 becomes the more relevant clustering criterion for describing the dynamics. \textbf{D:} The same 11 phenotypes now behave as a community of 3 effective species. Unlike a speciation event, where the uncertainty in defining the number of species is merely a matter of the desired level of detail (\textit{cf.} Fig.~\ref{fig:2}), the intermediate regime of panel C interpolates between unrelated clustering patterns (not nested).
\label{fig:3}}
\end{figure*}

The key property of the eigenmode spectrum of Fig.~\ref{fig:2}B is the existence of a noticeable \emph{gap} in its spectrum, which provides a justifiable cutoff point for the number of degrees of freedom to include into a low-dimensional description (note that in some contexts, the term ``spectral gap'' is used in a narrower sense to describe the difference between specifically the first and second eigenvalues; this is not the meaning adopted here). Whenever an ecosystem exhibits a spectral gap, one could define the effective number of species as the number of dominant eigenvalues. Importantly, however, the spectral gap need not be unique, and is a fluid property: under a continuous change of parameters, gaps can move, appear and disappear. The next example explores the implications of this fact.

\section{Moving the spectral gap:\\ ``two and a half species''?}
Consider now a community of 11 phenotypes of Fig.~\ref{fig:3}A. For equiabundant resources, we observe the expected ``2-species'' spectrum (Fig.~\ref{fig:3}B), the dominant eigenmodes reflecting the two ``core genomes'' that correspond to the dominating binary clustering pattern.

However, by construction (see SM for details), each of these phenotypes specializes in only one of the first three resources. If the supply of these resources is strongly increased, this preference will become the more relevant clustering criterion; the same 11 phenotypes now behave as a community of 3 effective species (Fig.~\ref{fig:3}D). Note that in this regime, the eigenmode corresponding to the binary clustering (which used to be mode \#2) is again part of the spectrum (now as mode \#4), but is characterized by a much smaller eigenvalue, signaling much slower dynamics than those governed by the dominant three resources.

Both these extremes were easy to interpret, but how should we think of the regime lying in between (Fig.~\ref{fig:3}C)?
A smooth process leading to a change in the number of species may seem familiar from the study of speciation. One cluster of phenotypes can be smoothly separated into two, and if we insisted on counting species, then somewhere in between there must be a gray area of ``more than one, but less than two''. This inability to draw a hard threshold of exactly when two close varieties should be considered distinct species is not, however, a significant challenge to the species paradigm. Whether a finer subdivision is warranted is merely a question of the desired level of detail. This was already illustrated in the previous example: the spectrum of Fig.~\ref{fig:2}B exhibits two significant gaps, leaving us the choice of characterizing it with either three or four degrees of freedom. And just as in Fig.~\ref{fig:3}, subjecting the community of Fig.~\ref{fig:2}A to an increased abundance of resources 1 and 2 would reduce the first spectral gap, making the finer (4-species) description increasingly more justifiable. In this scenario, the origin of our inability to define an exact number of species is well understood.

However, Fig.~\ref{fig:3} shows that the problem is more general. The somewhat artificial example of Fig.~\ref{fig:3}A was purposefully constructed to show that a smooth change of parameters can interpolate between two clustering patterns even when they are entirely unrelated (not nested): the same 11 phenotypes are grouped as \{1--5\}, \{6--11\} in one case (panel B), and \{1,~6\}, \{2,~7--10\}, \{3--5,~11\} in the other (panel D). The ``gray area'' in between is of a different nature than the uncertainty about whether to further subdivide a particular cluster. This suggests that the intermediate regime of Fig.~\ref{fig:3}C should be interpreted not as ``two and a half species'', but rather as a situation where a gap-less spectrum means that the effective number of species cannot be defined at all.

Characterizing such a ``regime without species'' is indeed our ultimate destination. Unfortunately, this particular example suffers from an obvious problem: it was constructed within a Lotka-Volterra framework with 11 degrees of freedom. Whether or not they can be grouped into clusters, such a model can always be seen as describing an interacting system of 11 species.

\begin{figure*}[t!]
\centering
\includegraphics[width=0.8\textwidth]{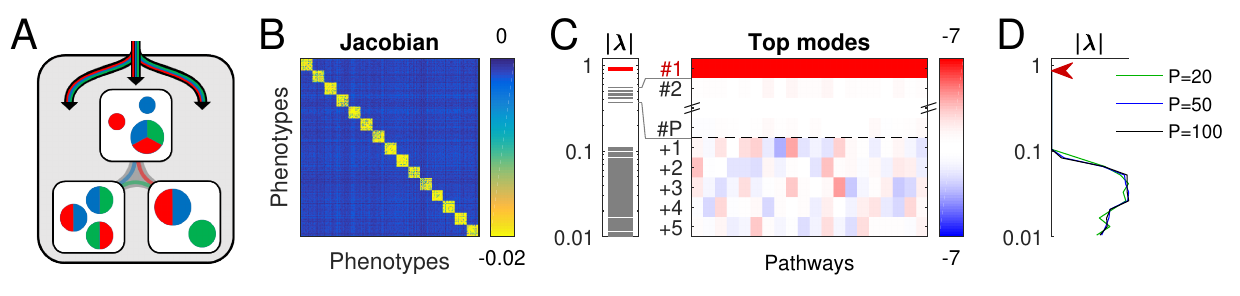}
\caption{Constructing a model case of an ecology without species.
\textbf{A:} A multi-patch extension of MacArthur's resource competition model. Externally supplied resources are split between $P$ patches harboring separate communities (here, $P=3$). Patch boundaries are assumed to be (weakly) permeable to resources, but not organisms (see SM for details). In this cartoon, phenotypes are represented by circles; size signifies abundance, while colors correspond to resources consumed.
\textbf{B:} The phenotype-phenotype interaction matrix for a particular realization of the model with $N=20$ resources and $P=15$ patches; the $298$ coexisting phenotypes are ordered by patch. Organisms in the same patch interact more strongly than across patches, leading to a block-diagonal structure.
\textbf{C:} The top $P$ eigenmodes reflect the $P$-patch structure of the habitat. Modes 2 through $P$ have vanishingly small projection into the pathway expression space: they represent inter-patch dynamics and are hidden from an external observer who cannot resolve individual patches.
\textbf{D:} The shape of the eigenvalue distribution is independent of $P$ (left); modes ``hidden'' from external observer (2 through $P$) are omitted. The unique dominant mode is the collective mode (arrowhead). The remainder of the spectrum follows a dense distribution with no gap, reflecting the unstructured background of interacting phenotypes.
\label{fig:4}}
\end{figure*}

This issue is a consequence of the decision, made early on, to work within the familiar Lotka-Volterra context. No simple Lotka-Volterra model can generate a truly gap-less spectrum: as discussed above, a model grouping individuals into $K$ discrete categories always has a gap separating the $K$ non-trivial eigenvalues from the vast number of those that are strictly zero, as a consequence of the partitioning assumption.
Similarly, \emph{virtually all classical ecological models describe systems whose eigenvalue spectrum exhibits a prominent gap.}
What reasons do we have to believe that natural ecosystems, e.g. the microbial communities inhabiting our intestines, possess this property?
Absent such evidence, we must acknowledge that spectral gaps, whose existence is a prerequisite for a low-dimensional ``species-based'' description to be adequate, could be an artifact of familiar theoretical models, and not necessarily the best approximation of ecological reality. If one could build an ecosystem model whose spectrum really did exhibit no gaps, such a scenario fundamentally could not be conceptualized as a system of interacting species. One signature of this regime is that any attempt to approximate it with a $K$-dimensional Lotka-Volterra model will appear to contain exactly $K$ species, for any $K$. The next section will construct one concrete example exhibiting this behavior.

\section{An ecology without species}
Constructing an ecosystem with a dense spectrum requires a model where the number of distinct phenotypes can be arbitrarily large. It would be convenient to also maintain two useful features of the previous examples: a functional characterization of phenotypes, and the existence of a stable equilibrium. Finally, to underscore the interpretation of this scenario as a regime ``without species'', we will set an additional objective, requiring the set of coexisting phenotypes to be unstructured, so that no grouping is privileged. Ideally, then, we would like to have a model where an arbitrarily large, unstructured set of functionally defined phenotypes can co-exist at a stable equilibrium.

In the classic (well-mixed) framework of MacArthur, $N$ resources can stably sustain at most $N$ phenotypes~\cite{MacArthur70}. However, imagine a spatially structured habitat with $P$ patches, whose boundaries are permeable to resources, but not to organisms (Fig.~\ref{fig:4}A). The mathematical details of this model are provided in the supplement; however, qualitatively, the behavior is intuitive. At infinite permeability (resources are exchanged freely), the $P$ patches are effectively a single patch, and we are back to the classic MacArthur model. If permeability is zero, the $P$ isolated patches can sustain up to $PN$ phenotypes, but these have no interactions across patch boundaries. At intermediate permeability, we find the desired situation where the interactions are nontrivial, and yet the total number of coexisting phenotypes can be arbitrarily large. Although the number of distinct binary vectors is bounded by $2^N$, two phenotypes located in different patches count as distinct even if their pathway content is identical, because they interact differently with other phenotypes (for example, of the two ``red-blue'' phenotypes in the cartoon in Fig.~\ref{fig:4}A, one has a strong competitor in the same patch, but the other does not).

For each patch, define an independent pool of local phenotypes that is fully random (each phenotype carries or not a given pathway with probability 1/2). After the system equilibrates, the total number of coexisting phenotypes is of order $PN$. Fig.~\ref{fig:4}B shows the complete interaction matrix for 298 phenotypes in $P=15$ patches (see SM for details). Predictably, phenotypes group by patch: at weak resource permeability, organisms interact more strongly within a patch than they do across patches. However, the pathway content of these phenotypes is, by construction, completely unstructured.

These observations are reflected in the eigenmode structure. $P$ eigenvalues clearly stand out in the spectrum (Fig.~\ref{fig:4}C), reflecting the patch structure: at the very top, we recognize the collective mode; the remaining $P-1$ modes are pairwise differences between the collective modes of the individual patches. These dominant modes are followed by a dense spectrum of eigenvalues, whose distribution exhibits no gap (Fig.~\ref{fig:4}D): aside from the patch structure, no grouping of phenotypes is in any way privileged, offering no possibility for a coarse-grained description.

In order to take the limit $P\rightarrow\infty$, consider now the projection of the eigenvectors into the $N$-dimensional expression space. This corresponds to adopting the perspective of an observer who can measure the community-wide expression of metabolic pathways, but cannot resolve individual patches. Such an observer can see the dominant collective mode, but modes 2 through $P$ are effectively hidden: their projection into the pathway expression space has vanishingly small components (panel~C). To understand this, consider a perturbation where all phenotypes in patch 1 experience an increase in abundance, and all phenotypes in patch 2, a decrease. Such a perturbation will relax quickly, corresponding to one of the dominant (strongly negative) eigenvalues. However, the internal dynamics represented by this mode is effectively hidden from an outside observer whose measurements do not resolve individual patches. Such an observer can only see a unique collective mode over a fully unstructured background (modes from $P+1$ onwards): a dense spectrum, whose shape is independent of $P$.

This characterization allows sending $P$ to infinity. This limit is proposed as a tractable model case of an ``ecology without species''. In this regime, the number of interacting phenotypes becomes infinite, their pathway content is unstructured, and the spectrum is dense; yet this construct continues to describe a valid ecology at equilibrium whose properties could be investigated. For instance, if the supply of some resource is increased, this system will have some well-defined response, which our hypothetical observer could measure and could attempt to predict.

What is the utility of this regime? The extreme, fully unstructured case is certainly a poor approximation of any natural setting. However, it provides a baseline that can be perturbed to describe novel scenarios of ``weakly structured'' ecologies, as illustrated in the next and final example.

\section{Describing weakly structured ecologies}
\begin{figure}[t!]
\centering
\includegraphics[width=\linewidth]{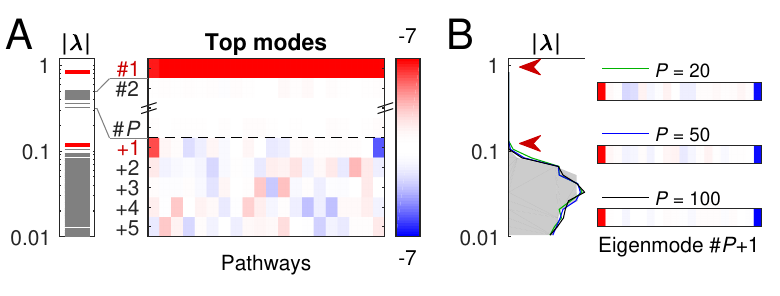}
\caption{A weakly structured ecology as a perturbation of the no-species regime. \textbf{A:}
When the phenotype pool is modified to introduce an anticorrelation between first and last pathway, the spectrum acquires a mode reflecting this structure (highlighted). This new mode is the dominant observable mode (after the collective mode). \textbf{C:} The leading observable eigenmodes remain at the same position in the spectrum (arrrowheads) and retain their structure (right). The rest of the spectrum follows the same distribution as in the unstructured regime (shaded; compare with Fig.~\ref{fig:4}D).
\label{fig:5}}
\end{figure}

Fig.~\ref{fig:5} repeats the analysis of Fig.~\ref{fig:4}, with one exception: the first and last pathway are forbidden from ever appearing within the same organism. Specifically, for each patch, the procedure generating the local phenotype pool was modified as follows: each entry in the pathway presence matrix $\sigma$ was independently set to 1 with probability 1/2 as before, but whenever this resulted in a phenotype carrying the forbidden pathway pair, this phenotype was removed from the pool. This kind of structure is not implausible: a detrimental cross-talk between certain pathways can make their simultaneous activity in the same organism difficult; for example, the enzyme nitrogenase is inactivated by oxygen and thus must be kept separate from oxygen-utilizing pathways.

The spectrum of this ecosystem, projected into the pathway expression space, is shown in Fig.~\ref{fig:5}A. Comparing to Fig.~\ref{fig:4}C, we see that there are now two modes that stand out from the gap-less background. In addition to mode \#1 (the collective mode), there is now also mode \#$(P+1)$ that represents the only perturbation imposed on an otherwise unstructured phenotype background. The spectrum in Fig.~\ref{fig:5}A was computed for one particular random realization of a system with $P=15$ patches; however, as $P$ is increased, the projection of this mode into pathway space and its position in the spectrum both remain invariant (Fig.~\ref{fig:5}B). This example demonstrates that the fully unstructured regime constructed here can indeed be meaningfully perturbed to describe the dynamics of ``weakly structured'' ecologies, an interesting theoretical scenario that was previously out of reach.

To maintain consistency with the previous examples, the approach taken here focused exclusively on resource competition. The ``hidden'' modes corresponding to unobservable inter-patch dynamics is the price to pay for this simplicity. Other approaches are possible; one might, for instance, imagine constructing a dense-spectrum model by means of a fully individual-based description, or through a sequence of iterative subdivisions of a Lotka-Volterra model (similar to the approach taken in Ref.~\cite{Kessler15}, but repeated an infinite number of times). One of these methods might yield a more elegant construction; however, the multi-patch example constructed here has the advantage of being easily interpretable, and potentially even analytically tractable with methods developed in Ref.~\cite{TikhonovMonasson16}.

\section{Discussion}
The mathematics of the ``eigenmode approach'' described here is very simple. Anyone familiar with the Principle Component Analysis (PCA) has encountered the problem of choosing the number of components on which to focus. In rare circumstances, a few components are sufficient to capture most of the variance; the generic situation, however, is that the spectrum shows no obvious gap, offering no natural truncation point~\cite{Bradde16}. Normally, one thinks of the PCA modes as linear combinations of the fundamental degrees of freedom, namely the species abundances. Here, the novelty is to ask whether the eigenmodes could be used as the fundamental degrees of freedom themselves.

Even if the concept of species is not called into question, the eigenmodes provide a naturally hierarchical description which allows asking new questions. Are there general laws governing the structure and evolution of community spectra? What ecological mechanisms might cause a given eigenvector to change in a predictable way, or a cause a given eigenvalue to become more prominent? How is the spectrum affected by evolutionary mechanisms? One may, for example, expect that purifying selection encourages the development of spectral gaps, while diversity-creating mutations have the opposite effect. A major advance in classical physics came with the development of the renormalization group, a framework explaining why many physical systems, as they become large, do in fact develop a gap in the spectrum, enabling some low-dimensional ``effective theory'' to become an excellent description~\cite{Machta13}. Ecology and evolution provide a new context for asking similar questions, which are particularly relevant today with the study of highly diverse microbial ecosystems of medical and environmental interest~\cite{HMP,EMP}.

The eigenmode formalism admits the possibility that, for a given ecosystem, the most relevant degrees of freedom may take a different form than simply the combined abundance of some taxa. As a simplest example, imagine a two-species consortium $S_1$, $S_2$ that runs the same metabolic reaction that species $S_3$ performs alone. In this scenario, the combination $S_1+S_2-S_3$ might be a functionally ``negligible'' (slow) eigenmode, but the reduction of dimensionality this observation affords does note reduce to simple branch-merging on a phylogenetic tree. Situations when one diverse set of species is essentially interchangeable with another diverse set are likely the rule, rather than an exception, as evidenced by the dramatic compositional variations of human-associated microbiota across healthy individuals, at all taxonomic levels~\cite{HMP}. This suggests that the number of negligible eigenmodes may indeed be large, and recognizing them could simplify the task of modelling these communities even within the traditional species-based framework.

However, more profoundly, the perspective developed here allows conceptualizing a novel scenario of an ``ecology without species''. The classic Lotka-Volterra model investigates an ecology of perfectly defined species, a regime that has been studied extensively. This work presented a theoretical construct that can be seen as the opposite extreme, one where the phenotypes are completely unstructured, and yet form a valid ecology.

In any real setting, organisms sharing a recent common ancestor are functionally similar, and the approximation ``every single organism is unlike any other'' is even more extreme than the converse assumption of perfect clustering into a small number of species. Nevertheless, a theoretical investigation of this unstructured regime will likely be a valuable step towards understanding the middle ground occupied by the real-life communities; in physical terms, its role would be to provide a different origin for a perturbative expansion (Fig.~\ref{fig:idea}). For example, in condensed matter physics it was recently proposed that a granular medium at the brink of a jamming transition can be thought of as a ``maximally disordered'' solid, the opposite extreme of a perfectly ordered crystal~\cite{Goodrich14}. Importantly, it was shown that certain properties of real-life solids, even at relatively low disorder, can be more adequately described as a perturbation of this ``maximally disordered'' regime, rather than as a perturbation of a perfect crystal~\cite{Goodrich14}. Similarly, it is intriguing to hypothesize that some behaviors of highly diverse natural ecosystems may be better described as a perturbation of a fully unstructured no-species regime (adding some tendency of phenotypes to cluster), rather than as a small deviation from the classic picture of well-clustered species (where we attempt to recognize that species boundaries are ``fuzzy''). This approach would provide a systematic framework to investigate the implications of ``structured'' and ``unstructured'' variation imposed at a range of scales, effects whose importance at the intraspecific level was recently highlighted~\cite{Moran16}.

The eigenmode formalism is by no means limited to resource competition models; MacArthur's model merely provided a convenient setting for constructing easily interpretable examples. The framework described here does, however, suffer from a number of other limitations: for example, only deterministic dynamics were considered, and the discussion was restricted to small fluctuations around some ecological state. Nevertheless, the question raised here was whether our description of ecosystems could move beyond the inherently discrete concept of a ``species''. In light of this challenge, the simplifications above provide a reasonable starting point.

\section{Acknowledgments}
I thank Ariel Amir, William Bialek, Michael P. Brenner, Carl P. Goodrich, Simon A. Levin, Anne Pringle, Ned S. Wingreen, and David Zwicker for helpful discussions. I have no competing interests. This work was supported by the Harvard Center of Mathematical Sciences and Applications and the Simons Foundation.

%\nolinenumbers

\onecolumngrid
 \appendix
 \cleardoublepage
 \section*{Supplementary material}
 \setcounter{figure}{0}
 \setcounter{equation}{0}
 \renewcommand{\theequation}{S\arabic{equation}}
 \renewcommand{\thefigure}{S\arabic{figure}}
 \renewcommand{\thetable}{S\arabic{table}}
\twocolumngrid

\section{The resource competition model (single patch)}
Examples in Figs.~\ref{fig:2} and~\ref{fig:3} were constructed within the resource competition framework used in Refs.~\cite{CWC,TikhonovMonasson}. For a detailed discussion of the metagenome partitioning model, its relation to the classic model of MacArthur~\cite{MacArthur69}, and the proof of existence, uniqueness and stability of the equilibrium state of its dynamics, the reader is referred to Ref.~\cite{CWC} (main text and supplementary material). The procedure for efficiently computing this equilibrium state numerically, as a convex $N$-dimensional optimization problem, is described in Ref.~\cite{TikhonovMonasson}. All computations were performed in MatLab (Mathworks, Inc.), and scripts reproducing all examples and figures are available as Supplementary File~1.

For the sake of completeness, the model will be briefly defined \emph{de novo}. The subsequent sections will describe the specifics of how examples in Figs.~\ref{fig:2} and~\ref{fig:3} were generated. The extension to a multi-patch habitat and Figs.~\ref{fig:4},~\ref{fig:5} will be discussed in a separate section.

\subsection{Definition of the model}
Consider a microbial community in a habitat where a single limiting element $\mathcal{X}$ (e.g.\ carbon) exists in $N$ forms (``resources'' $i\in\{1\dots N\}$) denoted $A$, $B$, etc. The substrates can be utilized with ``pathways'' $P_i$ (one specialized pathway per substrate). A phenotype $a$ is defined by the pathways that it carries. Specifically, its ``pathway content'' $\vsigma_a$ is a binary vector of pathway presence/absence: $\vec \sigma_a=\{\sigma_{ai}\}=\{1,1,0,1,\dots\}$.

For a given community, let $T_i$ be the total number of individuals capable of utilizing substrate $i$:
$$
T_i\equiv \sum_{a} n_a\sigma_{ai}.
$$
Here $n_a$ is the abundance of phenotype $a$ in the community (number of individuals with this phenotype). Assume a well-mixed environment, so that each of these $T_i$ individuals gets an equal share ${R_i}/{T_i}$ of the total benefit $R_i$ (e.g. carbon content) available from resource $i$. Any one resource is capable of sustaining growth, but accessing multiple cumulates the benefits. The fate of an individual of phenotype $a$ is determined by the \emph{resource surplus} $\Delta_a$ it experiences:
\begin{equation}\label{eq:surplus}
\Delta_a = \sum_i \sigma_{ai} \frac{R_i}{T_i} - \chi_a.
\end{equation}
Here the first term is the benefit harvested by all carried pathways, and $\chi_a$ represents the metabolic ``cost'' of phenotype $a$, namely its requirement for the growth-limiting element $\mathcal{X}$. The resource surplus $\Delta$ is ``spent'' on generating new biomass. Equating, as in Ref.~\cite{CWC}, the biomass of an organism with its cost for simplicity, we posit that for an individual of phenotype $a$, its division/death rate (as defined in the main text) is given by:
\begin{equation}\label{eq:dynamicsIndividual}
r_a = \frac{1}{\tau_0\chi_a} \,\Delta_a,
\end{equation}
where $\tau_0$ sets the unit of time. In terms of phenotype abundances, this dynamics translates into:
\begin{equation}\label{eq:dynamics}
\frac{dn_a}{dt} = g_a(\vec n) \equiv \frac{1}{\tau_0\chi_a}n_a\Delta_a.
\end{equation}
Here $\vec n$ is the vector of abundances of all phenotypes, and the function $g_a(\vec n)$ is defined by the right-hand side of this equation.

One does not \emph{need} to make the particular choice of equating phenotype biomass with its metabolic cost; other choices are possible. The same applies to the choice of resource competition as the unique ecological interaction on which to focus. Here and elsewhere, no claim is made that the particular choice made here is the best representation of biological reality. As explained in the main text, the examples used in this work were generated specifically to illustrate certain features of the eigenmode perspective, and most notably the role played by the gap in the model's spectrum. Many simplifying choices are made in order to make these illustrations easiest to interpret.

\subsection{Stability analysis}
The dynamics~\eqref{eq:dynamics} always lead to a unique, stable equilibrium~\cite{CWC,TikhonovMonasson}. This equilibrium may consist of fewer phenotypes, as some phenotypes may go extinct during equilibration. Denote $S$ the set of surviving phenotypes; for $a\in S$ the equilibrium condition reads $n_a>0$, $\Delta_a=0$. For the phenotypes that went extinct ($a\notin S$), we have $n_a=0$, $\Delta_a<0$.

We can now perform the stability analysis of the dynamics~\eqref{eq:dynamics} around the equilibrium point, i.e. consider the eigenmodes of the matrix
\begin{equation}\label{eq:M}
  M_{ab}=\frac{\partial g_a}{\partial n_b}.
\end{equation}
Considering separately the types that are present and those that are absent, one can write:
\begin{equation}\label{eq:jacobian}
M_{ab}=\left\{
\begin{aligned}
%\frac{\partial g_\alpha}{\partial n_\beta} &=
-&\frac {n_a}{\tau_0\chi_a} \sum_{i}\sigma_{ai} \sigma_{bi}\frac{R_i}{T_i^2} && \text{if $a\in S$}\\
%\frac{\partial g_\alpha}{\partial n_\beta} &=
&\frac 1 {\tau_0\chi_a}\delta_{ab}\Delta_a               && \text{if $a\notin S$}.
\end{aligned}
\right.
\end{equation}
Since the equilibrium is stable~\cite{CWC,TikhonovMonasson}, all eigenvalues of this Jacobian matrix are negative. Restricted to the space of phenotypes that are present, this matrix is given by:
\begin{equation}\label{eq:jacobianRestricted}
\left.M_{ab}\right|_{a,b\in S}=-\frac {n_a}{\tau_0\chi_a} \sum_{i}\sigma_{ai} \sigma_{bi}\frac{R_i}{T_i^2}
\end{equation}
Our object of interest is the eigensystem of this matrix.

\subsection{Lotka-Volterra vs. an individual-based description}
The main text considered two approaches to describing ecosystem dynamics: the discussion began with a general individual-based description (with an individual-individual interaction matrix $\MM$), which was then restricted to Lotka-Volterra-type dynamics for simplicity (with a phenotype-phenotype interaction matrix $M$). The main text claimed that if individuals fall into just $K$ categories, then the non-trivial eigenvalues of the individual-individual interaction matrix $\MM$ in the vicinity of equilibrium coincide with the eigenvalues of the $K$-by-$K$ matrix $M$ of the corresponding Lotka-Volterra model. (Assuming the abundances are sufficiently large that the continuous Lotka-Volterra model is a valid approximation.)

This claim is easy to verify. As above, let $n_a$ denote the abundances of the $K$ phenotypes, and let $M_{ab}$ be the matrix that defines linearized dynamics in the vicinity of this equilibrium point:
\begin{equation}\label{eq:contDynamics}
\frac{d\,\delta n_a}{dt} = \sum_{b=1}^K M_{ab} \,\delta n_b.
\end{equation}
Here $\delta n$ is the deviation from equilibrium abundance $n_a^{(0)}$. To convert this into an individual-based description, we ``split'' each phenotype into $n_a$ individuals, considered separately. What is the corresponding individual-individual interaction matrix?
For two individuals, $\mu$ of phenotype $a$ and $\nu$ of phenotype $b$, the interaction $M_{\mu\nu}$ is, by definition, the change in growth/death rate of $\nu$ that would be caused by a duplication of individual $\mu$. Let us make two observations: first, duplicating $\mu$ means increasing of $n_a$ by 1 unit. Second, the population growth rate of a phenotype $b$ and the division/death rate of an individual $\nu$ are related by the factor $n_\nu$. Therefore:
\begin{equation}\label{eq:MM}
M_{\mu\nu} = \frac1{n_b}M_{ab}.
\end{equation}
The final observation is that each eigenvector $v_a$ of the matrix $M_{ab}$ can be converted into an eigenvector of the matrix $\MM_{\mu\nu}$, by duplicating each entry $n_a$ times. The relation~\eqref{eq:MM} guarantees that the result will indeed be an eigenvector of $\MM$ with the same eigenvalue.

\subsection{Figure 2}
The purpose of the example shown in Fig.~2 was to illustrate that when phenotypes exhibit a clear structure, the eigenmode formalism successfully captures both the broad clustering pattern, and the more subtle subdivision. Accordingly, 4 ``core phenotypes'' were chosen, two of which are very close. The colors refer to the colored lines indicating the ``core phenotypes'' in Fig.~\ref{fig:2}:
\begin{enumerate}
\item carrying pathways 2 through 8 (green, variant 1);
\item carrying pathways 1 through 8, but not 2 (green, variant 2);
\item carrying pathways 9 through 14 (magenta).
\item carrying pathways 15 through 20 (cyan);
\end{enumerate}
The set of 20 competitors was constructed by taking five copies of each core phenotype, and randomly flipping pathway presence/absence with probability 0.05 to generate some diversity. Here, ``flipping a pathway'' means adding it to a phenotype where it was absent, or removing if it was present. The cost of each competitor was then assigned as follows:
$$
\chi_a = \sum_i \sigma_{ia}+\epsilon \xi_a.
$$
Here $\xi_a$ is a standard Gaussian random variable, and $\epsilon$ was set to 0.01. In other words, all competitors have the same cost per carried pathway, with a small perturbation. A cost that scales linearly with the number of carried pathways ensures that phenotypes carrying more pathways do not automatically have a competitive advantage~\cite{CWC}.

The community of 20 competitors that resulted from following the above procedure (for some particular seed of the pseudo-random number generator) was then equilibrated (with all resources supplied at equal rate, $R_i\equiv1$), following the numerical procedure described in Ref.~\cite{TikhonovMonasson}. As a result of this equilibration, some phenotypes went extinct; the 12 survivors are shown in Fig.~\ref{fig:2}.

\subsection{Figure 3}
The procedure generating this example is very similar: first, a pool of competitors with desired properties is constructed by hand. The competitors are assigned costs with a slight random component to avoid any accidental symmetries (e.g. two organisms whose costs are exactly the same). The community is equilibrated, and the pathway content of survivors is reported.

For Figure~3, phenotypes were constructed as variations on 6 ``core phenotypes''. Their structure (put in by hand) can be read off from Fig.~3A:
\begin{enumerate}
\item pathways 1 and 4--12;
\item pathways 2 and 4--12;
\item pathways 3 and 4--12;
\item pathways 1 and 13--21;
\item pathways 2 and 13--21;
\item pathways 3 and 13--21.
\end{enumerate}
Note that in this case $N=21$ pathways were used; this was done to ensure that all six core phenotypes carried the same number of pathways.

As before, a pool of 30 competitors was generated by taking 5 copies of each core phenotype, and adding/removing a few pathways to generate some diversity. To preserve the carefully prepared structure, pathways 1 through 3 were left intact, ensuring that all phenotypes specialize in exactly one of these first three resources. Of the remaining 18 pathways (numbered 4--21), exactly two were randomly ``flipped'' for each phenotype.

Finally, the pool of competitors was equilibrated for a range of resource conditions. The supply of resources 4 through 21 was fixed at 1, while the supply of resources 1, 2 and 3 was varied (synchronously) from $0.5$ to $50$. Only phenotypes that survived at non-zero abundance for the entire range of these conditions were retained. This procedure ensures that throughout the range of conditions shown in Fig.~3C, the community consists of the exact same set of phenotypes, in this case 11 of them.

At the end of this protocol, the resulting set of phenotypes was verified to contain at least one representative from each of the 6 core phenotypes. If this were not the case, the procedure would have been repeated for another choice of the seed for the pseudo-random generator.

\section{The multi-patch model: Figs~\ref{fig:4} and~\ref{fig:5}}
\subsection{Definition of the model}
In the single-patch model above, one can define the \emph{availability} of a resource as follows:
$$
h_i\equiv \frac{R_i}{T_i}.
$$
Recalling the definition of the ``resource surplus''
$$
\Delta_a = \sum_{i}\sigma_{ai}h_i-\chi_a,
$$
we see that $h_i$ determines the benefit that any organism receives from carrying the relevant pathway. Consider, for example, a specialist that only carries pathway \#1, and has cost $\chi_1$. Introduced in a community where the availability of resource 1 is $h_1$, this organism will be able to grow if and only if $\chi_1<h_1$. If this condition is satisfied, then this organism will multiply until it depletes resource 1 to exactly $h_1=\chi_1$. We see that $\chi_1$ plays the role of Tilman's $R^*$, and $h_1$ can be interpreted as the concentration of resource $i$ in the growth medium (up to some constant factor which ensures the right dimension). This argument is described in detail in Ref.~\cite{TikhonovMonasson}; the reason it is briefly reproduced here is to motivate the multi-patch construction that follows.

Consider two communities in two isolated patches; one depletes resource $i$ to $h_i^{(1)}$, the other to $h^{(2)}_i$. For concreteness, let $h^{(2)}_i<h^{(1)}_i$: the second community is more efficient at depleting resource $i$, and drives its equilibrium concentration to a lower value. If the patch boundary were permeable to this resource, we would expect a net flux of resource from the patch with higher concentration to a patch with lower concentration. In the simplest model, one can postulate that the magnitude of this flux (from the first patch into the second) is
$$
\delta R_i = \rho(h^{(1)}_i-h^{(2)}_i),
$$
where $\rho$ measures the permeability of patch boundaries. Let $\vec {\mathcal R}$ denote the global supply of resources (the vector notation is used to avoid explicitly specifying the index $i$, i.e. $\vec \RR\equiv \{\RR_i\}$). In isolation, both patches would receive an equal share of resource influx $\frac 12 \vec \RR$. However, the permeability of patch boundaries causes resources to be redistributed: the first patch now receives $\frac 12 \vec \RR-\vec{\delta R}$, and the second receives $\frac 12 \vec \RR+\vec{\delta R}$.

Generalizing this to $P$ patches, we obtain the following model for resource consumption in a multi-patch habitat. For each patch, the ecological dynamics remain identical to those of a single-patch MacArthur's model described above. The coupling between patches manifests itself only in the magnitudes of resource influx experienced by each patch. The global external supply $\vec \RR$ is evenly split between all $P$ patches; however, each patch may experience additional influx or outflux depending on the resource uptake efficiency of its local community.

Specifically, if $\alpha$ labels patches, define $\vec h^*$ as the \emph{average} resource availability across all patches:
$$
\vec h^* \equiv \frac 1P \sum_{\alpha=1}^P \vec h^{(\alpha)}.
$$
Patches where the availability (concentration) of resource $i$ is lower than $h^*_i$ experience a net influx of this resource from other patches, where the availability is higher. The total resource influx experienced by a given patch is therefore:
$$
\vec R^{(\alpha)} = \frac 1P \vec \RR - \vec {\delta R}^{(\alpha)},
$$
where
$$
\vec {\delta R}^{(\alpha)} = \rho(\vec h^{(\alpha)}_i-\vec h^*).
$$
At zero permeability ($\rho=0$), the patch dynamics are fully independent. At infinite permeability ($\rho\rightarrow\infty$), any difference in local resource availability would generate an infinitely strong flux, eliminating such difference; in this limit, the availability of resources in all patches must be identical $\vec h^{(\alpha)}\equiv \vec h^*$, and the standard single-patch MacArthur model is recovered. Both these limits have the expected behavior.

Numerically, for weak permeability $\rho$, the equilibrium of this dynamics can be found through a simple iterative procedure: each patch is separately equilibrated at fixed resource supply, at which point the resource exchange fluxes are computed, the amount of resource available at each patch is updated, and the procedure is repeated until convergence.

\subsection{Figs~\ref{fig:4} and~\ref{fig:5}}
To generate the example in Fig.~\ref{fig:4}, at each patch, a local pool of random 200 competitors for 20 resources was generated (at each patch, the pathway presence matrix was a random binary matrix of size 200-by-20, each entry set to 0 or 1 with probability 1/2). The number of patches was $P=15$ for the example shown in panels B and C, and increased to 20, 50 and 100 to compute the eigenvalue distributions in panel D. As before, the cost of each phenotype was set equal to the number of pathways it carried, with a small random contribution to break any accidental symmetries (adding a Gaussian random variable with zero mean and width $10^{-4}$). The permeability parameter was set to $1$ (a value empirically determined to have some weak effect on the set of coexisting phenotypes, inducing some interaction between patches), and the iterative equilibration procedure was performed until the relative magnitude of change in resource influx experienced by each patch would fall below $10^{-4}$.

To generate Fig.~\ref{fig:5}, the exact same procedure was followed, with one exception: prior to the equilibration procedure, any phenotype containing the ``forbidden'' pathway pair was removed from the competitor pool.

\end{document}